\documentclass[12pt,fleqn,thmsa]{article}
\usepackage{amsfonts}
\usepackage{amssymb,hyperref,backref}
\usepackage{amsmath,harvard,tikz,verbatim,epsfig}
\usepackage{color}
\usepackage{graphicx}

\newcommand{\be}{\begin{eqnarray*}}
\newcommand{\ee}{\end{eqnarray*}}

\newtheorem{theorem}{Theorem}[section]

{ 
}

\begin{document}

\title{{\large Dummy Endogenous Variables in Weakly
Separable \\ Multiple Index Models without Monotonicity}\thanks{
We are grateful to Jeremy Fox, Sukjin Han, Arthur Lewbel, Elie Tamer, Ed
Vytlacil, Haiqing Xu, and participants at the 2018 West Indies Economic
Conference, the 2019 SUFE econometrics meetings, and the 2020 Texas Camp
Econometrics for helpful comments and suggestions. }}
\author{{\small Songnian Chen} \\
{\small HKUST} \and {\small Shakeeb Khan} \\
{\small Boston College} \and {\small Xun Tang} \\
{\small Rice University }}
\date{{\footnotesize \today}}
\maketitle
\begin{center}  {\footnotesize {\bf Abstract} }\end{center}

\noindent {\footnotesize
We study the identification and estimation of treatment effect parameters in weakly separable models. 
In their seminal work, \citeasnoun{vytlacilyildiz} showed how to identify and estimate the average treatment effect of a dummy endogenous variable when the outcome is weakly separable in a single index. 
Their identification result builds on a monotonicity condition with respect to this single index.
In comparison, we consider similar weakly separable models with multiple indices, and relax the monotonicity condition for identification.
Unlike \citeasnoun{vytlacilyildiz}, we exploit the full information in the distribution of the outcome variable, instead of just its mean. 
Indeed, when the outcome distribution function is more informative than the mean, our method is applicable to more general settings than theirs; in particular we do not rely on their monotonicity assumption and at the same time we also allow for multiple indices. 
To illustrate the advantage of our approach, we provide examples of models where our approach can identify parameters of interest whereas existing methods would fail. 
These examples include models with multiple unobserved disturbance terms such as the Roy model and multinomial choice models with dummy endogenous variables, as well as potential outcome models with endogenous random coefficients. 
Our method is easy to implement and can be applied to a wide class of models. 
We establish standard asymptotic properties such as consistency and asymptotic normality.}

\noindent
{\small
} \pagebreak

\setcounter{equation}{0}

\section{Introduction}

\label{introduction}

Consider a weakly separable model with a binary endogenous variable:%
\begin{eqnarray}
Y &=&g(v_1(X,D),v_2(X,D),...v_J(X,D),\varepsilon ) \\
D &=&1\left \{ \theta (Z)-U>0\right \}
\end{eqnarray}%
where $( v_1(X,D),v_2(X,D),...v_J(X,D) ) \equiv v(X,D) $ is a $J$-vector of unknown linear or nonlinear indices in the outcome equation (1.1) and $D$ is a binary endogenous variable defined by the selection equation (1.2). 
Here $X\in \mathbb{R}^{d_{x}}$ and $Z\in \mathbb{R}%
^{d_{z}}$ are vectors of observable exogenous variables, which may have
overlapping elements. Similar to \citeasnoun{vytlacilyildiz} we require
exclusion restrictions that there is some element in $Z$ excluded from $X$,
and that we can vary $X$ after conditioning on $\theta(Z)$. In the system of equations above, $U$ is the unobservable random variable normalized to follow the uniform distribution $U(0,1)$ and the error term $\varepsilon$ in the outcome equation is allowed to be a random vector.  
  We
assume $\left( X,Z\right) $ are independent of $(\varepsilon ,U)$.  Note that we allow $v(X,D)$ to be a vector of multiple indices, whereas 
the method in \citeasnoun{vytlacilyildiz} can only be applied when it is a single index.

Since \citeasnoun{vytlacilyildiz}, other important work has considered
identification and estimation of related models, but under alternative
conditions. Examples with binary endogenous variables include %
\citeasnoun{hanvytlacil}, \citeasnoun{vuongxu}, \citeasnoun{lewbelqe}, %
\citeasnoun{khanetal}. Work for models when the endogenous variable is
continuous includes \citeasnoun{neweyimbens}, \citeasnoun{dhault2014} and %
\citeasnoun{torgo2014}. \citeasnoun{fengjun} shows how to identify
nonseparable triangular models where the endogenous variable is discrete and
has larger support than the instrument variable.\footnote{%
All these papers focus on point identification. For partial identification
of a model with a binary outcome, see \citeasnoun{shaikhvytlacil11} and %
\citeasnoun{mourifie15}.}

As in the conventional framework, two potential outcomes $%
Y_{1}$ and $Y_{0}$ satisfy%
\begin{equation*}
Y_D = g(v(X,D),\epsilon) \mbox{ for } D=0,1. \text{\ }
\end{equation*}%
We only observe $\left( Y,D,X,Z\right) $, where $Y=DY_{1}+(1-D)Y_{0}$. In
this model, as in \citeasnoun{vytlacilyildiz}, we do not impose parametric
distribution on the error term or a linear index structure. %
\citeasnoun{vytlacilyildiz} assumes that $v(X,D) \in \mathbb{R}$ is a single index, and
\begin{equation}
E\left[ g(v,\varepsilon )|U=u\right] \mbox{ is strictly increasing in }v \in \mathbb{R}%
\mbox{ for all }u.
\end{equation}%
Unlike \citeasnoun{vytlacilyildiz}, we do not impose any monotonicity
structure. That is because our approach exploits the full information in the
distribution of the outcome variable, instead of just its mean. Indeed, when
the outcome distribution function is more informative than the mean, our
method is applicable to more general settings than theirs; in particular we
do not rely on their monotonicity assumption and at the same time we also
allow for multiple indices.

In Sections \ref{Examples} and \ref{Extension} we provide some examples in
which such a monotonicity condition fails, but the average effect of the
binary endogenous variable is still identified. In addition, we allow for a
weakly separable model with multiple indices, that is, $v(X,D)=(v_{1}(X,D),v_{2}(X,D),...,v_{J}(X,D)) \in \mathbb{R}^J$.

We consider the identification and estimation of the average treatment effect of $D$ on $Y$, $E(Y_{1}|X\in A)$, $E(Y_{0}|X\in A)$ \ and $E(Y_{1}-Y_{0}|X\in A)$, for some set $A$, without the aforementioned monotonicity. 
Indeed, for the case with multiple indices $v(X,D) \in \mathbb{R}^J$, the monotonicity condition is no longer well defined.

\citeasnoun{vuongxu} established nonparametric identification of individual
treatment effects in a fully nonseparable model that includes a binary
endogenous regressor, without the nonlinear index structure. They assume $%
\varepsilon $ is a scalar and $g$ is strictly increasing in $\varepsilon $.
In their setting, monotonicity in the outcome equation provides the
identifying restriction to extrapolate information from local treatment
effects to population treatment effects.

\setcounter{equation}{0}

\section{Identification\label{Identification}}

Generally speaking, our identification strategy will be based on the notion
of \emph{matching}\footnote{%
See \citeasnoun{ahn-powell}, \citeasnoun{chenkhantang}, %
\citeasnoun{vytlacilyildiz}, and more recently \citeasnoun{auerbach} for
examples of papers that attain identification through matching.}. Consider
the identification of $E(Y_{1}|X=x)$ for some $x\in S_{1}$, where $S_{d}$
denotes the support of $X$ given $D=d\in \{0,1\}$. Note that because $%
(\varepsilon ,U)\bot (X,Z)$,%
\begin{eqnarray}
&&\text{ }E(Y_{1}|X=x)=E(Y_{1}|X=x,Z=z)  \notag \\
&&\text{ }=E(DY_{1}|X=x,Z=z)+E[(1-D)Y_{1}|X=x,Z=z]  \notag \\
&&\text{ }=P(z)E(Y|D=1,X=x,Z=z)+[1-P(z)]E(Y_{1}|D=0,X=x,Z=z)  \label{EQ1}
\end{eqnarray}%
where $P(z)\equiv E(D|Z=z)$. The only term that is not directly identifiable
on the right-hand side of (\ref{EQ1}) is%
\begin{equation*}
E(Y_{1}|D=0,X=x,Z=z)=E[g(v(x,1),\varepsilon )|U\geq P(z)]\text{.}
\end{equation*}%
The main idea behind our approach follows that of \citeasnoun{vytlacilyildiz}%
, which is to find some $\tilde{x}\in S_{0}\ $such that%
\begin{equation}
v(x,1)=v(\tilde{x},0)  \label{EQ2}
\end{equation}%
so that%
\begin{eqnarray*}
E(Y|D=0,X=\tilde{x},Z=z)= &&E(Y_{0}|D=0,X=\tilde{x},Z=z) \\
= &&E(g(v(\tilde{x},0),\varepsilon )|U\geq P(z))=E(g(v(x,1),\varepsilon
)|U\geq P(z))\text{.}
\end{eqnarray*}%
{Unlike \citeasnoun{vytlacilyildiz}, we utilize the full distribution of }$Y$
{\ (rather than its first moment) while searching for such pairs of }$(x,%
\tilde{x})${\ in (\ref{EQ2}). This allows us to relax the single-index and monotonicity
conditions in \citeasnoun{vytlacilyildiz}}.

For any $p$ on the support of $P(Z)$ given $X=x$, and for all $y$ define 
\begin{eqnarray}
h_{1}^{\ast }(x,y,p) &=&E(D1\left \{ Y\leq y\right \} |X=x,P(Z)=p)  \notag \\
&=&E\left[ 1\left \{ U<P(Z)\right \} 1\left \{ g(v(X,1),\varepsilon )\leq
y\right \} |X=x,P(Z)=p\right]  \notag \\
&=&\int_{0}^{p}F_{g|u}(y;v(x,1))du\text{,}  \label{EQ3}
\end{eqnarray}%
where 
\begin{equation*}
F_{g|u}(y;v(x,d))\equiv E[1\{g(v(x,d),\varepsilon )\leq y\}|U=u]\text{ }
\end{equation*}%
with $v(x,d)$ being a realized index at $X=x$ and the expectation in the
definition of $F_{g|u}$ is with respect to the distribution of $\varepsilon $
given $U=u$. The last equality in (\ref{EQ3}) holds because of independence
between $(\varepsilon ,U)$ and $(X,Z)$. By construction, $h_{1}^{\ast
}(x,y,p)$ is directly identified from the joint distribution of $(D,Y,X,Z)$
in the data-generating process. Furthermore, for any pair $p_{1}>p_{2}$,
define:%
\begin{equation*}
h_{1}(x,y,p_{1},p_{2})\equiv h_{1}^{\ast }(x,y,p_{1})-h_{1}^{\ast
}(x,y,p_{2})=\int_{p_{2}}^{p_{1}}F_{g|u}(y;v(x,1))du\text{.}
\end{equation*}%
Likewise, define 
\begin{eqnarray*}
h_{0}^{\ast }(x,y,p) &=&E((1-D)1\left \{ Y\leq y\right \} |X=x,P(Z)=p) \\
&=&E\left[ 1\left \{ U\geq P(Z)\right \} 1\left \{ g(v(X,0),\varepsilon )\leq
y\right \} |X=x,P(Z)=p\right] \\
&=&\int_{p}^{1}F_{g|u}(y;v(x,0))du.
\end{eqnarray*}%
and let 
\begin{equation*}
h_{0}(x,y,p_{1},p_{2})\equiv h_{0}^{\ast }(x,y,p_{2})-h_{0}^{\ast
}(x,y,p_{1})=\int_{p_{2}}^{p_{1}}F_{g|u}(y;v(x,0))du\text{.}
\end{equation*}%
{Let }$P_{x}${\ denote the support of }$P(Z)${\ given }$X=x${. It can be
shown that for any }$x\in S_{1}${\ and }$\tilde{x}\in S_{0}${,} and
any $y$,%
\begin{equation}
h_{1}(x,y,p,p^{\prime })=h_{0}(\tilde{x},y,p,p^{\prime })\text{ for all }%
p>p^{\prime }\text{ on }P_{x}\cap P_{\tilde{x}}\text{.}  \label{EQ4}
\end{equation}%
{if and only if}%
\begin{equation}
F_{g|p}(y;v(x,1))=F_{g|p}(y;v(\tilde{x},0))\text{ for all }p\in P_{x}\cap P_{%
\tilde{x}}\text{.}  \label{EQ5}
\end{equation}%
{Sufficiency is immediate from the definition of }$h_{1}${\ and }$h_{0}${.
To see necessity, note that for all }$p>p^{\prime }${\ on }$P_{x}\cap P_{%
\tilde{x}}${, }%
\begin{equation*}
\left. \frac{\partial h_{1}(x,y,\tilde{p},p^{\prime })}{\partial \tilde{p}}%
\right \vert _{\tilde{p}=p}=\left. \frac{\partial h_{1}^{\ast }(x,y,\tilde{p})%
}{\partial \tilde{p}}\right \vert _{\tilde{p}=p}=F_{g|p}(y;v(x_{1},1))
\end{equation*}%
{and}%
\begin{equation*}
\left. \frac{\partial h_{0}(\tilde{x},y,\tilde{p},p^{\prime })}{\partial 
\tilde{p}}\right \vert _{\tilde{p}=p}=-\left. \frac{\partial }{\partial 
\tilde{p}}h_{0}^{\ast }(\tilde{x},y,\tilde{p})\right \vert _{\tilde{p}%
=p}=F_{g|p}(y;v(\tilde{x},0))\text{.}
\end{equation*}%
{Thus (\ref{EQ4}) and (\ref{EQ5}) are equivalent.}

We collect the assumptions for identification as follows:\medskip

\noindent ASSUMPTION A-1: The distribution of $U$ is absolutely continuous
with respect to Lebesgue measure.\medskip

\noindent ASSUMPTION A-2: The random vectors $(U,\varepsilon )$ and $(X,Z)$
are independent.\medskip

\noindent ASSUMPTION A-3: The random variable $g(v(X,1),\varepsilon )$ and $%
g(v(X,0),\varepsilon )$ have finite first moments conditional on $U=u$ for
all $u \in [0,1]$..\medskip

\noindent ASSUMPTION A-4: For any $(x,\tilde{x})\in S_{1}\times S_{0}$, $%
F_{g|p}(y;v(x,1))=F_{g|p}(y;v(x,0))$ holds for all $y$ and $p\in P_{x}\cap
P_{\tilde{x}}$ if and only if $v(x,1)=v(\tilde{x},0)$.\medskip

\noindent ASSUMPTION A-5: $\Pr (X\in S_{1})>0$ \ and $\Pr (X\in S_{0})>0$%
.\medskip

Note that A-4 is weaker than Assumption 4 in \citeasnoun{vytlacilyildiz}.
Specifically, to identify pairs $(x,\tilde{x})$ with $v(x,1)=v(\tilde{x},0) $%
, \citeasnoun{vytlacilyildiz} relies on the assumption that $v(X,D) \in \mathbb{R}$ is a single index and that for any $%
(x,\tilde x)\in (S_1\times S_0)$, $E\left[ g(v(x,1),\varepsilon )|U=u\right]
=E\left[ g(v(\tilde{x},0),\varepsilon )|U=u\right] $ if and only if $%
v(x,1)=v(\tilde{x},0) $. There are two shortcomings with this approach.
First, it requires the condition (Assumption 4) that $E\left[
g(v(x,d),\varepsilon )|U=p\right] $ is a strictly monotonic function of $%
v(x,d)$. 
Second, when $v(x,d)$ is a vector of multiple indices instead of a single index, their approach breaks down. 
In comparison, we achieve the same purpose by matching
conditional distributions $F_{g|p}(\cdot ;v(x,1))$ and $\ F_{g|p}(\cdot ;v(%
\tilde{x},0))$. As we show in Section \ref{Examples}, in several important
applications, the outcome Y is either discrete (e.g. multinomial choices),
or multi-dimensional with both discrete and continuous components (e.g.,
potential outcomes determined by a Roy model). In either cases, the latent
index function $v(.)$ is vector-valued and the monotonicity condition in %
\citeasnoun{vytlacilyildiz} is not satisfied.

\setcounter{equation}{0}

\section{Examples}

\label{Examples}

{We now present several examples in which the latent indices are
multi-dimensional. In the first and third example, the monotonicity
condition in \citeasnoun{vytlacilyildiz} is not satisfied; in the second
example, the identification requires a generalization of the monotonicity
condition into an invertibility condition in higher dimensions.}\bigskip

\noindent \textbf{Example 1. (Heteroskaedastic shocks in outcome)} Consider
a triangular system where a continuous outcome is determined by double
indices $v(X,D)\equiv (v_{1}(X,D),v_{2}(X,D))$:%
\begin{equation*}
Y=g(v(X,D),\varepsilon )=v_{1}(X,D)+v_{2}(X,D)\varepsilon \text{ for }D\in
\{0,1\} \text{.}
\end{equation*}%
The selection equation determining the actual treatment is the same as
(1.2). In this case the concept of monotonicity in $v\in \mathbb{R}^{2}$ is
not well-defined, so the procedure proposed in \citeasnoun{vytlacilyildiz}
is not suitable here\footnote{%
For this particular design, the approach proposed in \citeasnoun{vuongxu}
should be valid. But it will not be for a slightly modified model, such as $%
Y=v_1(X,D)+(e_2+v_2(X,D)*e_1)$, whereas ours will be.}. Nevertheless, we can
apply the method in Section \ref{Identification}\ to identify the average
treatment effect by using the \textit{distribution} of outcome to find pairs
of $x$ and $\tilde{x}$ such that $v(x,1)=v(\tilde{x},0)$. Assume the range
of $v_2(\cdot)$ is positive. To see the necessity in Assumption A4, note that%
\begin{eqnarray*}
F_{g|u}(y;v(x,d)) &=&E\left[ v_{1}(x,d)+v_{2}(x,d)\varepsilon \leq y|U=u%
\right] \\
&=&F_{\varepsilon |u}\left( \frac{y-v_{1}(x,d)}{v_{2}(x,d)}\right)
\end{eqnarray*}%
for $d=0,1$. If the CDF of $\varepsilon $ is increasing over $\mathbb{R}$,
then for all $y$ and $x \in S_1$ and $\tilde{x} \in S_0$, 
\begin{equation*}
F_{g|u}(y;v(x,1))=F_{g|u}(y;v(\tilde{x},0))
\end{equation*}%
if and only if%
\begin{equation*}
\frac{y-v_{1}(x,1)}{v_{2}(x,1)}=\frac{y-v_{1}(\tilde{x},0)}{v_{2}(\tilde{x}%
,0)}\text{.}
\end{equation*}%
Differentiating with respect to $y$ yields%
\begin{equation*}
v_{2}(x,1)=v_{2}(\tilde{x},0)
\end{equation*}%
which in turn implies%
\begin{equation*}
v_{1}(x,1)=v_{1}(\tilde{x},0)\text{.}
\end{equation*}%
The sufficiency in Assumption A-4 is straight-forward.\bigskip

\noindent \textbf{Example 2. (Multinomial potential outcome)} Consider a
triangular system where the outcome is multinomial. The multinomial response
model has a long and rich history in both applied and theoretical
econometrics. Recent examples in the semiparametric literature include %
\citeasnoun{lee95}, \citeasnoun{ahnetal2018}, \citeasnoun{shietal2018}, %
\citeasnoun{pakesporter2014}, \citeasnoun{KOT2019}. But unlike the work
here, none of those papers allow for dummy endogenous variables or potential
outcomes.

\begin{equation*}
Y=g(v(X,D),\varepsilon )=\arg \max_{j=0,1,...,J}y_{j,D}^{\ast }
\end{equation*}%
where 
\begin{equation*}
y_{j,D}^{\ast }=v_{j}(X,D)+\varepsilon _{j}\text{ } \mbox{ for } j =
1,2,...,J\text{; }y_{0,D}^{\ast }=0\text{.}
\end{equation*}%
In this case the index $v\equiv (v_{j})_{j\leq J}$ and the errors $%
\varepsilon \equiv (\varepsilon _{j})_{j\leq J}$ are both $J$-dimensional.
The selection equation that determines $D$ is the same as (1.2). In this
case, we can replace $1\{Y\leq y\}$ by $1\{Y=y\}$ in the definition of $%
h_{1},h_{0},h_{1}^{\ast },h_{0}^{\ast }$ and $F_{g|u}(\cdot ;v)$. Then for $%
d=0,1$ and $j\leq J$,\ 
\begin{eqnarray*}
F_{g|u}(j;v(x,d)) &\equiv &E[1\{g(v(x,d),\varepsilon )=j\}|U=u] \\
&=&\Pr \left \{ v_{j}(x,d)+\varepsilon _{j}\geq v_{j^{\prime
}}(x,d)+\varepsilon _{j^{\prime }}\text{ }\forall j^{\prime }\leq J\mid
U=u\right \} \text{.}
\end{eqnarray*}%
By \citeasnoun{ruud2000} and \citeasnoun{ahnetal2018}, the mapping from $%
v\in \mathbb{R}^{J}$ to $(F_{g|u}(j;v):j\leq J)\in \mathbb{R}^{J}$ is smooth
and invertible provided that $\varepsilon \in \mathbb{R}^{J}$ has
non-negative density everywhere. This implies Assumption A-4.\bigskip

\noindent \textbf{Example 3}. \textbf{(Potential outcome from the Roy model)}
Consider a treatment effect model with an endogenous binary treatment $D$
and with the potential outcome determined by a latent Roy model. The Roy
model has also been studied extensively from both applied and theoretical
perspectives. See for example the literature survey in %
\citeasnoun{heckvyt-handbook} and the seminal paper in %
\citeasnoun{heckmanhonoreb}.

Here the observed outcome consists of two pieces:\  \ a continuous measure $%
Y=DY_{1}+(1-D)Y_{0}$ and a discrete indicator $W=DW_{1}+(1-D)W_{0}$ for $%
d=0,1$. These potential outcomes are given by 
\begin{equation*}
Y_{d}=\max_{j\in \{a,b\}}y_{j,d}^{\ast }\text{ and }W_{d}=\arg \max_{j\in
\{a,b\}}y_{j,d}^{\ast }
\end{equation*}%
where $a$ and $b$ index potential outcomes realized in different sectors,
with 
\begin{equation*}
y_{j,d}^{\ast }=v_{j}(X,d)+\varepsilon _{j}\text{.}
\end{equation*}%
The binary endogenous treatment $D$ is determined as in the selection
equation (1.2). For example, $D\in \{1,0\}$ indicates whether an individual
participates in certain professional training program, $W_{d}\in \{a,b\}$
indicates the potential sector in which the individual is employed, $%
y^*_{j,d}$ is the potential wage from sector $j$ under treatment $D=d$, and $%
Y_{d}\in \mathbb{R}$ is the potential wage if the treatment status is $D=d$.
As before, we maintain that $(X,Z)\bot (\varepsilon ,U)$.

The parameter of interest is 
\begin{equation*}
\Pr \{Y_{1}\leq y,W_{1}=a|X\}
\end{equation*}%
which by the independence $(X,Z)\bot (\varepsilon ,U)$ and an application of
the law of total probability can be decomposed into directly identifiable
quantities and a counterfactual quantity 
\begin{eqnarray}
&&\Pr \{Y_{1}\leq y,W_{1}=a\mid X=x,Z=z,D=0\}  \notag \\
&=&\Pr \left \{ v_{b}(x,1)+\varepsilon _{b}<v_{a}(x,1)+\varepsilon _{a}\leq
y\mid U\geq P(z)\right \} \text{.}  \label{CF}
\end{eqnarray}%
Again, we seek to identify this counterfactual quantity by finding $\tilde{x}%
\in S_{0}$ such that 
\begin{equation}
v_{a}(x,1)=v_{a}(\tilde{x},0)\text{ and }v_{b}(x,1)=v_{b}(\tilde{x},0)\text{ 
}  \label{fullEQ}
\end{equation}%
This would allow us to recover the right hand side of (\ref{CF}) as 
\begin{equation*}
\Pr \{Y_{0}\leq y,W_{0}=a\mid X=\tilde{x},Z=z,D=0\} \text{.}
\end{equation*}

To find such a pair of $(x,\tilde{x})$, define $h_{d,W}(x,p,p^{\prime
}),h_{d,W}^{\ast }(x,p)$ by replacing $1\{Y\leq y\}$ with $1\{W=a\}$ in the
definition of $h_{d},h_{d}^{\ast }$ in Section \ref{Identification}.
Similarly, define $h_{d,Y}(x,y,p,p^{\prime }),h_{d,Y}^{\ast }(x,y,p)$ by
replacing $1\{Y\leq y\}$ with $1\{Y\leq y,W=a\}$ in the definition of $%
h_{d},h_{d}^{\ast }$ in Section \ref{Identification}. Then 
\begin{eqnarray*}
h_{d,W}(x,p_{1},p_{2}) &=&\int_{p_{2}}^{p_{1}}\Pr \{v_{b}(x,d)+\varepsilon
_{b}<v_{a}(x,d)+\varepsilon _{a}|U=u\}du\text{;} \\
h_{d,Y}(x,y,p_{1},p_{2}) &=&\int_{p_{2}}^{p_{1}}\Pr \{v_{b}(x,d)+\varepsilon
_{b}<v_{a}(x,d)+\varepsilon _{a}\leq y|U=u\}du\text{;}
\end{eqnarray*}%
and $h_{d,W}(x,p_{1},p_{2})$ and $h_{d,Y}(x,y,p_{1},p_{2})$ are both
identified over their respective domains by construction. Assume $%
(\varepsilon _{a},\varepsilon _{b})$ is continuously distributed with
positive density over $\mathbb{R}^{2}$ conditional on all $u$. Then the
statement 
\begin{eqnarray*}
\text{\textquotedblleft }h_{1,W}(x,p,p^{\prime }) &=&h_{0,W}(\tilde{x}%
,p,p^{\prime })\text{ \textit{and} }h_{1,Y}(x,y,p,p^{\prime })=h_{0,Y}(%
\tilde{x},y,p,p^{\prime })\text{ } \\
\text{for all }y\text{ and }p &>&p^{\prime }\text{ on }P_{x}\cap P_{\tilde{x}%
}\text{\textquotedblright }
\end{eqnarray*}%
holds true if and only if (\ref{fullEQ}) holds. Then matching $%
h_{1,W}(x,p,p^{\prime })=h_{0,W}(\tilde{x},p,p^{\prime })$ ensures 
\begin{equation}
v_{a}(x,1)-v_{b}(x,1)=v_{a}(\tilde{x},0)-v_{b}(\tilde{x},0)\text{;}
\label{CF1}
\end{equation}
while matching $h_{1,Y}(x,y,p,p^{\prime })=h_{0,Y}(\tilde{x},y,p,p^{\prime })
$ \textit{at the same time} ensures that in addition to (\ref{CF1}) 
\begin{equation}
v_{a}(x,1)=v_{a}(\tilde{x},0)\text{.}  \label{CF2}
\end{equation}%
Combining (\ref{CF1}) and (\ref{CF2}) is equivalent to (\ref{fullEQ}).

\setcounter{equation}{0}

\section{Extension}

\label{Extension}

{The identification strategy we have used so far requires matching exogenous
variables }$x,\tilde{x}${\ on }$S_{0},S_{1}$. In some cases, with the
outcome being continuous, we can construct similar argument for identifying
a counterfactual quantity in a treatment effect model by matching different
elements on the support of continuous outcome. This approach was not
investigated in \citeasnoun{vytlacilyildiz}, which focused on the use of
first moment of outcome. The following example illustrates this
point.\bigskip

\noindent \textbf{Example 4.} \textbf{(Potential outcome with random
coefficients)} Random coefficient models are prominent in both the
theoretical and applied econometrics literature. They permit a flexible way
to allow for conditional heteroscedasticity and unobserved heterogeneity.
See, for example \citeasnoun{hsiaochapter} for a survey. Here we consider a
treatment effect model where the potential outcome is determined through
random coefficients: 
\begin{equation*}
Y=DY_{1}+(1-D)Y_{0}\text{ where }Y_{d}=(\alpha _{d}+X^{\prime }\beta _{d})%
\text{ for }d=0,1
\end{equation*}%
and the binary endogenous treatment $D$ is determined as in the selection
equation (1.2). The \textit{random} intercepts $\alpha _{d}\in \mathbb{R}$
and the \textit{random} vectors of coefficients $\beta _{d}$ are given by 
\begin{equation*}
\alpha _{d}=\bar{\alpha}_{d}(X)+\eta _{d}\text{ and }\beta _{d}=\bar{\beta}%
_{d}(X)+\varepsilon _{d}
\end{equation*}%
where for any $x$ and $d \in \{0,1\}$., $(\bar{\alpha}_{d}(x),\bar{\beta}%
_{d}(x))\in \mathbb{R}^{K+1}$ is a vector of constant parameters while $\eta
_{d}\in \mathbb{R}$ and $\varepsilon _{d}\in \mathbb{R}^{K}$ are
unobservable noises.

As in \citeasnoun{vytlacilyildiz}, assume some elements in $Z$ in the
selection equation are excluded from $X$. We allow the vector of
unobservable terms $(\epsilon _{1},\epsilon _{0},\eta _{0},\eta _{1},U)$ to
be arbitrarily correlated. We also assume that%
\begin{equation}
\left( X,Z\right) \text{ }\bot \text{ }(\epsilon _{1},\epsilon _{0},\eta
_{0},\eta _{1},U)\text{,}  \label{indep}
\end{equation}%
with the marginal distribution of $U$ normalized to standard uniform, so
that $\theta (Z)$ is directly identified as $P(Z)\equiv E(D|Z=z)$.

In this example our goal is to identify the conditional distribution of $%
Y_{d}$ given $X=x$ for $d=0,1$. From this result we can identify parameters
of interest such as average treatment effects, quantile treatment effects,
etc. Let $G_{P|x}$ denote the conditional distribution of $P\equiv P(Z)$
given $X=x$, which is directly identifiable from the data-generating
process. By construction,%
\begin{equation*}
\Pr \{Y_{1}\leq y|X=x\}=\int \Pr \{Y_{1}\leq y|X=x,P=p\}dG_{P|x}(p)\text{,}
\end{equation*}%
where%
\begin{eqnarray*}
&&\Pr \{Y_{1}\leq y|X=x,P=p\} \\
&=&E\left[ D1\{Y_{1}\leq y\}|X=x,P=p\right] +E\left[ (1-D)1\{Y_{1}\leq
y\}|X=x,P=p\right] \text{.}
\end{eqnarray*}%
The first term on the right-hand side is identified as%
\begin{equation*}
E[D1\{Y\leq y\}|X=x,P=p]\text{,}
\end{equation*}%
while the second term is counterfactual and can be written as%
\begin{eqnarray*}
&&\phi _{0}(x,y,p)\equiv E[1\{U\geq P\}1\{ \alpha _{1}+X^{\prime }\beta
_{1}\leq y\}|X=x,P=p] \\
&=&E[1\{U\geq p\}1\{ \bar{\alpha}_{1}(x)+\eta _{1}+x^{\prime }(\bar{\beta}%
_{1}(x)+\varepsilon _{1})\leq y\}] \\
&=&\int_{p}^{1}\Pr \{ \eta _{1}+x^{\prime }\epsilon _{1}\leq y-\bar{\alpha}%
_{1}(x)-x^{\prime }\bar{\beta}_{1}(x)|U=u\}du\text{.}
\end{eqnarray*}%
For any $p$ on the support of $P$ given $X=x$, define%
\begin{eqnarray*}
&&h_{1}^{\ast }(x,y,p)\equiv E\left[ D1\left \{ Y\leq y\right \} |X=x,P=p%
\right] \\
&=&E\left[ 1\left \{ U<P\right \} 1\left \{ \alpha _{1}+X^{\prime }\beta
_{1}\leq y\right \} |X=x,P=p\right] =E\left[ 1\left \{ U<p\right \} 1\left \{
\alpha _{1}+x^{\prime }\beta _{1}\leq y\right \} \right] \\
&=&\int_{0}^{p}\Pr \{ \eta _{1}+x^{\prime }\epsilon _{1}\leq y-\bar{\alpha}%
_{1}(x)-x^{\prime }\bar{\beta}_{1}(x)|U=u\}du,
\end{eqnarray*}%
where the second equality uses (\ref{indep}). Likewise, under (\ref{indep})
we have: 
\begin{eqnarray*}
&&h_{0}^{\ast }(x,y,p)\equiv E\left[ (1-D)1\left \{ Y\leq y\right \} |X=x,P=p%
\right] \\
&=&\int_{p}^{1}\Pr \{ \eta _{0}+x^{\prime }\epsilon _{0}\leq y-\bar{\alpha}%
_{0}(x)-x^{\prime }\bar{\beta}_{0}(x)|U_{i}=u\}du\text{.}
\end{eqnarray*}%
Assume\footnote{%
This type of distributional equality assumption generalizes the exact
equality of $\epsilon_1,\epsilon_0$ as can be found in for example %
\citeasnoun{vytlacilyildiz}. Distributional equality has been used to
motivate the \emph{rank similarity} condition imposed frequently in the
econometrics literature- see for example \citeasnoun{chernozhukovhansen}, %
\citeasnoun{franlef}, \citeasnoun{dongshen}, \citeasnoun{chenstackhan}.}

\begin{equation}
F_{\left( \eta _{1},\epsilon _{1}\right) |U=u}=F_{\left( \eta _{0},\epsilon
_{0}\right) |U=u}\text{ for all }u\in \lbrack 0,1]\text{.}  \label{eqCF}
\end{equation}%
Under (\ref{eqCF}), we have%
\begin{equation}
\phi _{0}(x,y,p)=\int_{p}^{1}\Pr \{ \eta _{0}+x^{\prime }\epsilon _{0}\leq y-%
\bar{\alpha}_{1}(x)-x^{\prime }\bar{\beta}_{1}(x)|U=u\}du\text{.}
\label{swapEps}
\end{equation}%
Suppose for each pair $(x,y)$ we can find $t(x,y)$ such that 
\begin{equation*}
y-\bar{\alpha}_{1}(x)-x^{\prime }\bar{\beta}_{1}(x)=t(x,y)-\bar{\alpha}%
_{0}(x)-x^{\prime }\bar{\beta}_{0}(x)\text{.}
\end{equation*}%
Then by construction%
\begin{eqnarray*}
h_{0}^{\ast }(x,t(x,y),p) &\equiv &\int_{p}^{1}\Pr \{ \eta _{0}+x^{\prime
}\epsilon _{0}\leq t(x,y)-\bar{\alpha}_{0}(x)-x^{\prime }\bar{\beta}%
_{0}(x)|U=u\}du \\
&=&\int_{p}^{1}\Pr \{ \eta _{0}+x^{\prime }\epsilon _{0}\leq y-\bar{\alpha}%
_{1}(x)-x^{\prime }\bar{\beta}_{1}(x)|U=u\}du=\phi _{0}(x,y,p)
\end{eqnarray*}%
because of (\ref{swapEps}). Thus the counterfactual $\phi _{0}(x,y,p)$ would
be identified as $h_{0}^{\ast }(x,t(x,y),p)$.

It remains to show that for each pair $(x,y)$ we can uniquely recover $%
t(x,y) $ using quantities that are identifiable in the data-generating
process. To do so, we define two auxiliary functions as follows: for $%
p_{1}>p_{2}$ on the support of $P$ given $X=x$, let 
\begin{eqnarray*}
h_{1}(x,y,p_{1},p_{2}) &\equiv &h_{1}^{\ast }(x,y,p_{1})-h_{1}^{\ast
}(x,y,p_{2}) \\
&=&\int_{p_{2}}^{p_{1}}\Pr \{ \eta _{1}+x^{\prime }\epsilon _{1}<y-\bar{\alpha%
}_{1}(x)-x^{\prime }\bar{\beta}_{1}(x)|U=u\}du\text{;}
\end{eqnarray*}%
and%
\begin{eqnarray*}
h_{0}(x,y,p_{1},p_{2}) &\equiv &h_{0}^{\ast }(x,y,p_{2})-h_{0}^{\ast
}(x,y,p_{1}) \\
&=&\int_{p_{2}}^{p_{1}}\Pr \{ \eta _{0}+x^{\prime }\epsilon _{0}<y-\bar{\alpha%
}_{0}(x)-x^{\prime }\bar{\beta}_{0}(x)|U=u\}du\text{.}
\end{eqnarray*}%
Suppose $\eta _{d}+x^{\prime }\epsilon _{d}$ is continuously distributed
over $\mathbb{R}$ for all values of $x$ conditional on all $u \in [0,1]$.
Then for any fixed pair $(x,y)$ and $p_{1}<p_{2}$,%
\begin{equation*}
h_{1}(x,y,p_{1},p_{2})=h_{0}(x,t(x,y),p_{1},p_{2})
\end{equation*}%
if and only if%
\begin{equation*}
t(x,y)=y-\bar{\alpha}_{1}(x)-x^{\prime }\bar{\beta}_{1}(x)+\bar{\alpha}%
_{0}(x)+x^{\prime }\bar{\beta}_{0}(x)\text{.}
\end{equation*}%
To see this, suppose $t(x,y)>y-\bar{\alpha}_{1}(x)-x^{\prime }\bar{\beta}%
_{1}(x)+\bar{\alpha}_{0}(x)+x^{\prime }\bar{\beta}_{0}(x)$, then (\ref{eqCF}%
) implies that $h_{0}(x,t(x,y),p_{1},p_{2})>h_{1}(x,y,p_{1},p_{2})$. A
symmetric argument establishes a similar statement with \textquotedblleft $>$%
\textquotedblright \ replaced by \textquotedblleft $<$\textquotedblright .
This establishes our desired result. 

\setcounter{equation}{0}

\section{Estimation}

\label{estimation}

Here we outline estimation procedures from a random sample of the observed
variables that are motivated by our identification results. We will first
describe an estimation procedure for the parameter $E[Y_1]$ in the first
three examples. Let $P_{x}$ to denote the support of $P(Z)$ given $X=x$, $%
f_p(.|x)$ denote the density of $P(Z)$ given $X=x$, and 
\begin{equation*}
P_{x}^{c}=\left \{ p\text{: }f_{p}(p|x)>c\right \} \mbox{ for a known }c>0,
\end{equation*}%
and for simplicity assume a strong overlap condition that%
\begin{equation*}
1-c_{0}>P(Z)>c_{0} \mbox{ for a known } c_0>0,
\end{equation*}%
Define a measure of distance between $h_{1}(x_{1},\cdot )$ and $%
h_{0}(x_{0},\cdot )$ 
\begin{eqnarray*}
&&\left \Vert h_{1}(x_{1},\cdot )-h_{0}(x_{0},\cdot )\right \Vert \\
&=&\left \{ \int \int \int \left( \int_{p_{1}}^{p_{2}} (F_{g|u}(y;v(x_1,1)) -
F_{g|u}(y;v(x_0,0)) du \right)^{2} I \left( p_{1},p_{2}\in P_{x}^{c}\right)
w(y)dydp_{1}dp_{2}\right \}^{1/2}
\end{eqnarray*}
where $w(y)$ is a chosen weight function. Consider the case when $%
h_{1}(x,y,p_{1},p_{2})$, $h_{1}(x,y,p_{1},p_{2})$ and $P(z)$ are known. For
any given $X_{i}$, let $\tilde X_{i}$ be such that%
\begin{equation*}
\left \Vert h_{1}(X_{i},\cdot )-h_{0}(\tilde X_{i},\cdot )\right \Vert =0
\end{equation*}
which, under Assumption A-4 in Section \ref{Identification}, is equivalent
to 
\begin{equation*}
v(X_{i},1)=v(\tilde X_{i},0).
\end{equation*}%
Define%
\begin{equation*}
\hat{Y}_{i}=E(Y|D=0, \| h_1(X_i,\cdot)- h_0( X,\cdot)\|=0,P=P_{i}).
\end{equation*}%
Note that the conditional expectation on the right-hand side is equal to
\linebreak $E[ Y | D=0, v( X , 0 ) = v( X_i , 1), P=P_i] $, which in turn
equals $E[ Y_1 | D=0, X=X_i, P=P_i] $. Then, following the discussion above,
we define the following estimator for $\Delta \equiv E[Y_{1}]$: 
\begin{equation*}
\hat{\Delta}=\frac{1}{n}\sum_{i=1}^{n}\left( D_{i}Y_{i}+(1-D_{i})\hat{Y}%
_{i}\right)
\end{equation*}%
or a \textbf{weighted version}%
\begin{equation*}
\hat{\Delta}_{w}=\frac{\frac{1}{n}\sum_{i=1}^{n}1\left \{ X_{i}\in A\right \}
\left( D_{i}Y_{i}+(1-D_{i})\hat{Y}_{i}\right) }{\frac{1}{n}%
\sum_{i=1}^{n}1\left \{ X_{i}\in A\right \} }
\end{equation*}%
Limiting distribution theory for each of these estimators follows from 
identical arguments in \citeasnoun{vytlacilyildiz}. Here we formally state
the theorem for the first estimator:

\begin{theorem}
\label{theorem1} Under Assumptions A-1 to A-5, and the additional assumption
that $Y_1$ has positive and finite second moment, then we have 
\begin{equation*}
\sqrt{n}(\hat \Delta-\Delta)\overset{d}{\rightarrow} \mathbb{N}(0,V)
\end{equation*}
where 
\begin{equation*}
V=Var(E[Y_1|X,P,D])+E[PVar(Y_1|X,P,D=1)]
\end{equation*}
\end{theorem}

Now we describe an estimation procedure for the distributional treatment
effect in Example 4, where we had a model with random coefficients. In this
case, the parameter of interest is for a chosen value of the scalar $y$, 
\begin{equation*}
\Delta_2(y)=\Pr \{Y_{1}\leq y\} \text{.}
\end{equation*}%
First, for fixed values of $y$ and $p_{1}>p_{2}$, we propose to estimate $%
t(x,y)$ as 
\begin{equation*}
\hat{t}(x,y,p_{1},p_{2})=\arg
\min_{t}(h_{1}(x,y,p_{1},p_{2})-h_{0}(x,t,p_{1},p_{2}))^{2}
\end{equation*}%
and then average over values of $p_{1},p_{2}$: 
\begin{equation*}
\hat{\tau}(x,y)=\frac{1}{n(n-1)}\sum_{i\neq j}I[P_{i}>P_{j}]\hat{t}%
(x,y,P_{i},P_{j})
\end{equation*}%
An infeasible estimator for the parameter $\Delta_2(y)$,
which assumes $t(x,y)$ is known, would be%
\begin{equation*}
\hat{\Delta}_2(y)=\frac{1}{n}\sum_{i=1}^{n} \left( D_{i}1\{Y_{i}\leq
y\}+(1-D_{i})1\{Y_{i}\leq t(X_{i},y)\} \right) \text{.}
\end{equation*}%
In practice, for feasible estimation, one needs to replace $t(x,y)$ by its
estimator $\hat{\tau}(x,y)$.


\setcounter{equation}{0}

\section{Simulation Study}

\label{simulations}

This section presents simulation evidence for the performance of the
proposed estimation procedures described in Section \ref{estimation}, for
both the Average Treatment Effect and the Distributional Treatment Effect.
We report results for both our proposed estimator and that in %
\citeasnoun{vytlacilyildiz}, for several designs. These include designs
where the said monotonicity condition fails, and designs where the
disturbance terms in the outcome equation are multidimensional.

Throughout all designs we model the treatment or dummy endogenous variable
as 
\begin{equation*}
D=I[Z-U>0]
\end{equation*}
where $Z,U$ are independent standard normal. We experiment with the
following designs for the outcome

\begin{description}
\item[Design 1] 

\begin{equation*}
Y=X+0.5\cdot D +\epsilon
\end{equation*}
where $X$ is standard normal, $(\epsilon, U)$ are distributed bivariate
normal, each with mean 0 and variance 1, with correlations of 0,0.25,0.5.


\item[Design 2] 
\begin{equation*}
Y=X+0.5\cdot D+(X+D)\cdot \epsilon
\end{equation*}
where $X$ is distributed standard normal, $(\epsilon, U)$ are distributed
bivariate normal, each with mean 0 and variance 1, with correlations of
0,0.25,0.5.

\item[Design 3] 
\begin{equation*}
Y=(X+0.5 \cdot D+\epsilon)^2
\end{equation*}
where $X$ is distributed standard normal, $(\epsilon, U)$ are distributed
bivariate normal, each with mean 0 and variance 1, with correlations of
0,0.25,0.5.
\end{description}

We note that the monotonicity condition is satisfied in design 1  but fails
in the other two designs. For each of these designs, we report results for
estimating the parameter $E[Y_1]$, which denotes the expected value for
potential outcome under treatment $D=1$. The two estimators used in the
simulation study were the one proposed in Section \ref{estimation} and the
method proposed in \citeasnoun{vytlacilyildiz}. The summary statistics,
scaled by the true parameter value, Mean Bias, Median Bias, Root Mean
Squared Error, (RMSE), and Median Absolute Deviation (MAD) were evaluated
for sample sizes of 100, 200, 400 for 401 replications. Results for each of these designs are reported 
in Tables 1 to 3 respectively. In implementing  our procedure we assumed the propensity
score function is known, and conducted next stage estimation using a
nonparametric kernel estimator with normal kernel function, and a bandwidth
of $n^{-1/5}$. This rate reflects ``undersmoothing" as there are two
regressors, the propensity score and the regressor $X$. For the estimator in %
\citeasnoun{vytlacilyildiz}, which involved the derivative of conditional
expectation functions as well, estimating these functions nonparametrically
gave very unstable results so we report results for an infeasible version of
their estimator, assuming such functions, as well as the propensity scores,
are known.

To implement the second stage of our proposed procedure, in calculating the
distance $\|h_1(x_i,\cdot)-h_0(x_0,\cdot)\|$ we used an evenly space grid of
values for $y$, and selected $n/50$ grid points, with $n $ denoting the
sample size. 

The results indicate the desirable properties of our proposed procedure,
generally agreeing with Theorem \ref{theorem1}. In all designs our estimator
has small values for bias and RMSE, with the value of RMSE decreasing as the
sample size grows. In contrast, the procedure based on %
\citeasnoun{vytlacilyildiz} only performs well in Design 1, with values of
bias and RMSE comparable to those using our method. As in our procedure
these values decrease with as the sample size grows, which is expected, as
the monotonicity condition rely on is satisfied in these designs. In this
case, their approach has smaller standard errors largely due to the relative
simpler structure of the infeasible version, but their biases persist even
when the sample size increases.

For designs 2 and 3, where monotonicity is violated, the procedure proposed
in \citeasnoun{vytlacilyildiz} does not perform well. In design 2 in Table 2
both the bias and RMSE are generally increasing with the sample size.
Results for their estimator are better in design 3, but the bias hardly
converges with the sample size and is much larger compared to our estimator.

We also simulate data from a model with dummy endogenous variable and
potential outcomes determined by random coefficients. It is important to
note that for this design, the original matching idea in %
\citeasnoun{vytlacilyildiz} does not apply. This is because different values
of $x$ lead to different distribution of the composite error $\eta_d +
x^{\prime }\epsilon_d$. Our contribution in Section \ref{Extension} is to
propose a new approach based on matching different values of outcome $y$,
rather than the regressors $x$. Based on the counterfactual framework
discussed in Section \ref{Extension}, here the treatment variable $D$ is
modeled as the same way as the dummy endogenous variable above. Similarly
the regressor $X$ is standard normal. For both $Y_0,Y_1$ the random
intercepts were modeled as constants (0 and 1, respectively) and the
additive error terms were each standard normal. For the random slopes, the
means were 1 and 2 respectively, and the additive error terms were also
standard normal, independent of all other disturbance terms and each other.
Here we use the procedure in Section \ref{Extension} to estimate the
parameter $\Delta_2=P(Y_1<y)$, where in the simulation we set $y=1$. The
same four summary statistics are reported for sample sizes 100,200,400,
based on 401 replications. 
Results for this random coefficients design are reported in Table 4.

The estimator proposed in Section \ref{estimation} performs well; but the
bias and RMSE are much small at 400 observations compared to 100 and 200
observations, indicating convergence at the parametric rate. \pagebreak


\begin{center}
{\tiny 
}

{\tiny {\ Table 1} }

{\tiny 
\begin{tabular}{||lccc|ccc||}
\hline
\  & \  & CKT &  &  & VY &  \\ \hline \hline
$\rho _{v}$ & $0$ & $1/4$ & $1/2$ & $0$ & $1/4$ & $1/2$ \\ \hline
n=100 & \  & \  & \  & \  & \  &  \\ 
MEAN BIAS & -0.0170 & 0.0229 & -0.0435 & -0.1302 & -0.1676 & -0.2018 \\ 
MEDIAN BIAS & -0.0137 & 0.0124 & -0.0653 & -0.1318 & -0.1678 & -0.2087 \\ 
RMSE & 0.4936 & 0.4800 & 0.4945 & 0.3308 & 0.3337 & 0.3546 \\ 
MAD & 0.3289 & 0.3328 & 0.3156 & 0.2200 & 0.2271 & 0.2546 \\ 
n=200 & \  & \  & \  & \  & \  &  \\ 
MEAN BIAS & 0.0032 & -0.0024 & -0.0069 & -0.0864 & -0.1299 & -0.1766 \\ 
MEDIAN BIAS & -0.0102 & -0.0141 & -0.0314 & -0.0934 & -0.1277 & -0.1679 \\ 
RMSE & 0.3355 & 0.3367 & 0.3521 & 0.2293 & 0.2457 & 0.2711 \\ 
MAD & 0.2240 & 0.2228 & 0.2517 & 0.1594 & 0.1676 & 0.1865 \\ 
n=400 & \  & \  & \  & \  & \  &  \\ 
MEAN BIAS & -0.0187 & 0.0101 & -0.0055 & -0.0584 & -0.11134 & -0.1593 \\ 
MEDIAN BIAS & -0.0261 & 0.0128 & -0.0065 & -0.0592 & -0.1162 & -0.1572 \\ 
RMSE & 0.2496 & 0.2489 & 0.2578 & 0.2049 & 0.1867 & 0.2167 \\ 
MAD & 0.1523 & 0.1732 & 0.1659 & 0.1197 & 0.1345 & 0.1605 \\ \hline \hline
\end{tabular}
}

{\tiny {\ Table 2} }

{\tiny 
\begin{tabular}{||lccc|ccc||}
\hline
\  & \  & CKT &  &  & VY &  \\ \hline \hline
$\rho _{v}$ & $0$ & $1/4$ & $1/2$ & $0$ & $1/4$ & $1/2$ \\ \hline
n=100 & \  & \  & \  & \  & \  &  \\ 
MEAN BIAS & 0.0109 & 0.0397 & -0.0671 & -0.1509 & -0.2875 & -0.4207 \\ 
MEDIAN BIAS & 0.0151 & 0.0227 & -0.0939 & -0.1590 & -0.2918 & -0.4262 \\ 
RMSE & 0.5089 & 0.2737 & 0.4853 & 0.3524 & 0.4199 & 0.5289 \\ 
MAD & 0.3395 & 0.2447 & 0.3105 & 0.2419 & 0.30898 & 0.4310 \\ 
n=200 & \  & \  & \  & \  & \  &  \\ 
MEAN BIAS & 0.0322 & 0.0143 & -0.0311 & -0.1273 & -0.2559 & -0.3875 \\ 
MEDIAN BIAS & 0.0159 & 0.0054 & -0.0543 & -0.1310 & -0.2553 & -0.3884 \\ 
RMSE & 0.3487 & 0.3444 & 0.3455 & 0.2622 & 0.3407 & 0.4475 \\ 
MAD & 0.2317 & 0.2297 & 0.2552 & 0.1782 & 0.2624 & 0.3884 \\ 
n=400 & \  & \  & \  & \  & \  &  \\ 
MEAN BIAS & 0.0088 & 0.0269 & -0.0294 & -0.0962 & -0.2247 & -0.3708 \\ 
MEDIAN BIAS & 0.0007 & 0.0244 & -0.0309 & -0.0982 & -0.2255 & -0.3769 \\ 
RMSE & 0.2578 & 0.2557 & 0.2549 & 0.1920 & 0.2764 & 0.4037 \\ 
MAD & 0.1649 & 0.1733 & 0.1606 & 0.1354 & 0.2283 & 0.3769 \\ \hline \hline
\end{tabular}
} 

{\tiny {\ Table 3} }

{\tiny 
\begin{tabular}{||lccc|ccc||}
\hline
\  & \  & CKT &  &  & VY &  \\ \hline \hline
$\rho _{v}$ & $0$ & $1/4$ & $1/2$ & $0$ & $1/4$ & $1/2$ \\ \hline
n=100 & \  & \  & \  & \  & \  &  \\ 
MEAN BIAS & -0.0097 & -0.0070 & 0.0019 & -0.0691 & -0.0898 & -0.1066 \\ 
MEDIAN BIAS & -0.0233 & -0.0101 & -0.0240 & -0.0799 & -0.0925 & -0.1178 \\ 
RMSE & 0.1893 & 0.2085 & 0.2126 & 0.1546 & 0.1630 & 0.1701 \\ 
MAD & 0.1398 & 0.1342 & 0.1374 & 0.1125 & 0.1178 & 0.1315 \\ 
n=200 & \  & \  & \  & \  & \  &  \\ 
MEAN BIAS & -0.0108 & -0.0069 & -0.0068 & -0.0609 & -0.0765 & -0.0968 \\ 
MEDIAN BIAS & -0.0148 & -0.0033 & -0.0099 & -0.0674 & -0.0769 & -0.1017 \\ 
RMSE & 0.1372 & 0.1434 & 0.1424 & 0.1163 & 0.1262 & 0.1369 \\ 
MAD & 0.0949 & 0.0989 & 0.0953 & 0.0855 & 0.0887 & 0.1078 \\ 
n=400 & \  & \  & \  & \  & \  &  \\ 
MEAN BIAS & -0.0073 & -0.0014 & -0.0026 & -0.0583 & -0.0725 & -0.0889 \\ 
MEDIAN BIAS & -0.0149 & -0.0023 & -0.0029 & -0.0610 & -0.0751 & -0.0887 \\ 
RMSE & 0.1084 & 0.0994 & 0.0989 & 0.0924 & 0.1007 & 0.1131 \\ 
MAD & 0.0697 & 0.0685 & 0.0654 & 0.0689 & 0.0788 & 0.0901 \\ \hline \hline
\end{tabular}
}

\pagebreak {\tiny {\ Table 4} }

{\tiny 
\begin{tabular}{||lccc||}
\hline
\  & CKT &  &  \\ \hline \hline
$\rho _{v}$ & $0$ & $1/4$ & $1/2$ \\ \hline
n=100 & \  & \  &  \\ 
MEAN BIAS & 0.0109 & -0.0086 & 0.0038 \\ 
MEDIAN BIAS & 0.0000 & -0.0064 & 0.0126 \\ 
RMSE & 0.1011 & 0.0979 & 0.0955 \\ 
MAD & 0.0600 & 0.0648 & 0.0652 \\ 
n=200 & \  & \  & \  \\ 
MEAN BIAS & -0.0050 & -0.0150 & 0.0095 \\ 
MEDIAN BIAS & -0.0100 & -0.0161 & 0.0029 \\ 
RMSE & 0.0669 & 0.0669 & 0.0665 \\ 
MAD & 0.0400 & 0.0454 & 0.0457 \\ 
n=400 & \  & \  & \  \\ 
MEAN BIAS & 0.0012 & -0.0132 & 0.0074 \\ 
MEDIAN BIAS & 0.0049 & -0.0162 & 0.0077 \\ 
RMSE & 0.0501 & 0.0494 & 0.0495 \\ 
MAD & 0.0349 & 0.0325 & 0.0360 \\ \hline \hline
\end{tabular}
}
\end{center}

{\normalsize \setcounter{equation}{0} }

\section{Conclusion}

{\normalsize In this paper, we considered identification and estimation of 
nonseparable models with endogenous binary treatment. Existing approaches
are based on a monotonicity condition, which is violated in models with
multiple unobserved idiosyncratic shocks. Such models arise in many
important empirical settings, including Roy models and multinomial choice
models with dummy endogenous variables, as well as treatment effect models
with random coefficients. We establish novel identification results for
these models which are constructive and conducive to estimation procedures
which are easy to compute and whose limiting distributional properties
follow from standard large sample theorems. A simulation study indicates
adequate finite sample performance of our proposed methods. }

{\normalsize This paper leaves open areas for future research. Our method
requires the selection of the number and location of cutoff points, so a
data driven method for selecting these would be useful. Furthermore, the
relative efficiency of our proposed approach needs to be explored, perhaps
by deriving efficiency bounds for these new classes of models. }

{\normalsize 
\bibliographystyle{econometrica}
\bibliography{overall}
}

\end{document}